\newcommand{\SiIIa}{Si~{\sc ii} $\lambda 6355$}

\newcommand{\msol}{\mbox{M$_{\odot}$}}

\newcommand{\kms}{\mbox{$\rm{km}\,s^{-1}$}}

\newcommand{\ps}{\protect \hbox {Pan-STARRS1}}
\newcommand{\obj}{SN~2022ilv}
\newcommand{\hvf}{SN~2020hvf}

\documentclass[twocolumn]{aastex631}

\shorttitle{SN 2022ilv}
\shortauthors{Srivastav et al.}
\graphicspath{{./}{figures/}}

\begin{document}

\title{The luminous type Ia supernova 2022ilv and its early excess emission}

\correspondingauthor{Shubham Srivastav}
\email{s.srivastav@qub.ac.uk}

\author[0000-0003-4524-6883]{Shubham Srivastav}
\affiliation{Astrophysics Research Centre, School of Mathematics and Physics, Queen’s University Belfast, Belfast BT7 1NN, UK}
\author[0000-0002-8229-1731]{S. J. Smartt}
\affiliation{Astrophysics Research Centre, School of Mathematics and Physics, Queen’s University Belfast, Belfast BT7 1NN, UK}
\affiliation{Department of Physics, University of Oxford, Keble Road, Oxford OX1 3RH}
\author[0000-0003-1059-9603]{M. E. Huber}
\affiliation{Institute for Astronomy, University of Hawaii, 2680 Woodlawn Drive, Honolulu, HI 96822, USA}
\author[0000-0001-9494-179X]{G. Dimitriadis}
\affiliation{School of Physics, Trinity College Dublin, The University of Dublin, Dublin 2, Ireland}
\author[0000-0001-6965-7789]{K. C. Chambers}
\affiliation{Institute for Astronomy, University of Hawaii, 2680 Woodlawn Drive, Honolulu, HI 96822, USA}
\author[0000-0003-1916-0664]{Michael D. Fulton}
\affiliation{Astrophysics Research Centre, School of Mathematics and Physics, Queen’s University Belfast, Belfast BT7 1NN, UK}
\author[0000-0001-8385-3727]{Thomas Moore}
\affiliation{Astrophysics Research Centre, School of Mathematics and Physics, Queen’s University Belfast, Belfast BT7 1NN, UK}
\author[0000-0002-7975-8185]{F. P. Callan}
\affiliation{Astrophysics Research Centre, School of Mathematics and Physics, Queen’s University Belfast, Belfast BT7 1NN, UK}
\author[0000-0002-8094-6108]{James H. Gillanders}
\affiliation{Department of Physics, University of Rome ``Tor Vergata'', Via della Ricerca Scientifica 1, I-00133 Rome, Italy}
\author[0000-0002-9770-3508]{K. Maguire}
\affiliation{School of Physics, Trinity College Dublin, The University of Dublin, Dublin 2, Ireland}
\author[0000-0002-2555-3192]{M. Nicholl}
\affiliation{Birmingham Institute for Gravitational Wave Astronomy and School of Physics and Astronomy, University of Birmingham, Birmingham B15 2TT, UK}
\author[0000-0002-5738-1612]{Luke J. Shingles}
\affiliation{GSI Helmholtzzentrum f\"{u}r Schwerionenforschung, Planckstraße 1, 64291 Darmstadt, Germany}
\affiliation{Astrophysics Research Centre, School of Mathematics and Physics, Queen’s University Belfast, Belfast BT7 1NN, UK}
\author[0000-0002-9774-1192]{S. A. Sim}
\affiliation{Astrophysics Research Centre, School of Mathematics and Physics, Queen’s University Belfast, Belfast BT7 1NN, UK}
\author[0000-0001-9535-3199]{K. W. Smith}
\affiliation{Astrophysics Research Centre, School of Mathematics and Physics, Queen’s University Belfast, Belfast BT7 1NN, UK}
\author[0000-0003-0227-3451]{J. P. Anderson}
\affiliation{European Southern Observatory, Alonso de C\'ordova 3107, Casilla 19, Santiago, Chile}
\affiliation{Millennium Institute of Astrophysics MAS, Nuncio Monsenor Sotero Sanz 100, Off.
104, Providencia, Santiago, Chile}
\author[0000-0001-5486-2747]{Thomas de Boer}
\affiliation{Institute for Astronomy, University of Hawaii, 2680 Woodlawn Drive, Honolulu, HI 96822, USA}
\author[0000-0002-1066-6098]{Ting-Wan Chen}
\affiliation{Technische Universit{\"a}t M{\"u}nchen, TUM School of Natural Sciences, Physik-Department, James-Franck-Stra{\ss}e 1, 85748 Garching, Germany}
\affiliation{Max-Planck-Institut f{\"u}r Astrophysik, Karl-Schwarzschild Stra{\ss}e 1, 85748 Garching, Germany}
\author[0000-0003-1015-5367]{Hua Gao}
\affiliation{Institute for Astronomy, University of Hawaii, 2680 Woodlawn Drive, Honolulu, HI 96822, USA}
\author[0000-0002-1229-2499]{D. R. Young}
\affiliation{Astrophysics Research Centre, School of Mathematics and Physics, Queen’s University Belfast, Belfast BT7 1NN, UK}


\begin{abstract}

We present  observations and analysis of the host-less and luminous type Ia supernova 2022ilv, illustrating it is part of the 2003fg-like family, often referred to as super-Chandrasekhar (Ia-SC) explosions. The ATLAS light curve shows evidence of a short-lived, pulse-like early excess, similar to that detected in another luminous type Ia supernova (\hvf). The light curve is broad and the early spectra are remarkably similar to  SN 2009dc. Adopting a redshift of $z=0.026 \pm 0.005$ for SN 2022ilv based on spectral matching, our model light curve requires a large  $^{56}$Ni mass in the range $0.7-1.5$ \msol, and a large ejecta mass in the range $1.6-2.3$ \msol.
The early excess can be explained by fast-moving SN ejecta interacting with a thin, dense shell of circumstellar material close to the progenitor ($\sim 10^{13}$~cm), a few hours after the explosion. This may be realised in a double-degenerate scenario, wherein a white dwarf merger is preceded by ejection of a small amount ($\sim 10^{-3}-10^{-2}$ \msol) of hydrogen and helium-poor tidally stripped material. A deep pre-explosion Pan-STARRS1 stack indicates no host galaxy to a limiting magnitude of $r \sim 24.5$. This implies a surprisingly faint limit for any host of $M_r \gtrsim -11$, providing further evidence that these types of explosion occur predominantly in low-metallicity environments. 

\end{abstract}



\section{Introduction} \label{sec:intro}

A rare subclass of type Ia supernovae (SNe Ia) are often referred to as `super-Chandrasekhar' Ia or Ia-SC. The 
observational properties of this subclass are defined by the prototype SN 2003fg \citep{2006Natur.443..308H} 
along with SNe 2006gz, 2007if, 2009dc \citep{2007ApJ...669L..17H,2010ApJ...713.1073S,2011MNRAS.412.2735T} among others. 
Their broad light curves would require $\gtrsim 1$ \msol\ of radioactive $^{56}$Ni and well over a Chandrasekhar mass ($M_{\rm Ch}$) of ejecta. However, we now understand that this is a diverse subclass of objects \citep{2021ApJ...922..205A}, with peak luminosities ranging from $M_B^{\rm peak} \approx -19.1$ for ASASSN-15hy \citep{2021ApJ...920..107L}, to $M_B^{\rm peak} \approx -20.4$ for SN 2006gz and SN 2009dc \citep{2010ApJ...713.1073S,2011MNRAS.412.2735T}. 

The subclass is generally characterised by high luminosities, broad light curves, relatively low expansion velocities, and high ultraviolet (UV) and Near Infrared (NIR) luminosities, posing a challenge for theoretical models \citep{2017hsn..book..317T}. Most of these events do not follow the width-luminosity relation, typically showing negative Hubble residuals \citep{2021ApJ...922..205A}. 

Early observations at a few hours to days past explosion are a vital diagnostic tool for investigating the SN Ia explosion mechanism and nature of the binary companion \citep{2014ARA&A..52..107M}. A significant fraction ($\sim 20\%$) show evidence of a flux excess in their early light curves \citep{2020A&A...634A..37M,2022MNRAS.512.1317D}. Multiple physical mechanisms have been proposed -- SN ejecta colliding with a binary companion within the single-degenerate scenario \citep[e.g.][]{2010ApJ...708.1025K}, SN ejecta interacting with extended circumstellar material  \citep[e.g.][]{2017MNRAS.470.2510L}, presence of $^{56}$Ni clumps in the outermost layers of the ejecta \citep[e.g.][]{2019ApJ...870...13S}, and helium shell detonations occurring on the WD surface in the double-detonation scenario \citep[e.g.][]{2019ApJ...873...84P}.

Recently, an early flux excess was detected for the peculiar, luminous Ia \hvf\ by \citet{2021ApJ...923L...8J}, who invoked circumstellar material (CSM) interaction involving $\sim 10^{-2}$ \msol\ of extended material at a distance of $\sim 10^{13}$ cm. A tentative early excess was also reported for ASASSN-15pz \citep{2019ApJ...880...35C} and LSQ12gpw \citep{2018ApJ...865..149J}. It is possible that the early excess is ubiquitous in this subclass which may help understand their progenitor channel. In this paper, we present another clear discovery of a flux excess in the early light curve of a luminous type Ia (\obj) with extensive photometric, spectroscopic data and light curve models.

\section{Discovery and Follow-up}\label{sec:disc}

\obj\ was discovered by the Zwicky Transient Facility  
\citep[ZTF;][]{2019PASP..131a8002B} on 2022 April 24.30 UT, or MJD 59693.30, at a magnitude of $r=18.32\pm0.07$.
It was classified by \citet{2022TNSCR1137....1B} on MJD 59698.63 as a super-Chandrasekhar SN Ia (Ia-SC) at a redshift of $z=0.031$. 

We independently flagged it as a `real' transient in the Asteroid Terrestrial-impact Last Alert System \citep[ATLAS;][]{2018PASP..130f4505T} survey data on MJD 59692.48, a day prior to ZTF discovery \citep{2020PASP..132h5002S}. The  ATLAS units covered the position of \obj\ on 4 epochs (each with 4$\times$30s) between 59689.39 and 59992.48, all before the ZTF discovery  (see Section~\ref{subsec:excess}). 

We triggered followup photometric and spectroscopic observations of \obj\ at the 2m Liverpool Telescope \citep[LT;][]{2004SPIE.5489..679S} under programs PL22A20 and PL22B16 (PI Srivastav), and PL22A13 (PI Dimitriadis). Point-spread function (PSF) photometry in $griz$ was calibrated against Pan-STARRS1 (PS1) reference stars and the $u$-band was calibrated using Sloan Digital Sky Survey (SDSS) data. 
ZTF magnitudes in $gr$ bands were obtained through the Lasair 
broker\footnote{https://lasair-ztf.lsst.ac.uk/object/ZTF22aahhywm/}
\citep{2019RNAAS...3a..26S}. We also obtained images in $iz$ bands using the 1.8m Pan-STARRS2 (PS2) telescope \citep{Chambers2016}. The PS2 data were processed through the Image Processing Pipeline \citep[IPP;][]{2020ApJS..251....3M} and image subtraction was performed using PS1 $3\pi$ survey data \citep{Chambers2016} as reference.

\obj\ was also observed by the Ultraviolet and Optical Telescope \citep[UVOT;][]{2005SSRv..120...95R} on  Swift  \citep{2004ApJ...611.1005G}. Four epochs of  imaging  were obtained and photometry was performed using the \texttt{uvotsource} task within the High Energy Astrophysics SOFTware (\textsc{heasoft}) package, following \citet{2009AJ....137.4517B}. 
 
The photometry of \obj\ and \hvf\ is summarised in Table~\ref{tab:phot_2022ilv}.

\begin{table}
\centering
\caption{Summary of photometric observations of \obj\ and \hvf. All magnitudes are in the AB system.}
\begin{tabular}{ccccccc}
\hline
SN & MJD & Mag & Error & Instrument & Filter\\
\hline
2022ilv & 59689.38 & 19.36 & 0.28 & ATLAS & o\\
2022ilv & 59690.52 & $>20.02$ & $-$ & ATLAS & o\\
2022ilv & 59691.26 & 19.44 & 0.15 & ATLAS & o\\
2022ilv & 59692.48 & 18.64 & 0.09 & ATLAS & c\\
2022ilv & 59694.49 & 17.70 & 0.03 & ATLAS & o\\
2022ilv & 59695.21 & 17.36 & 0.02 & ATLAS & o\\
\hline
\end{tabular}
\label{tab:phot_2022ilv}
\tablecomments{This table is published in its entirety in the machine-readable format. A part of the table is shown here for guidance regarding its form and content.}
\end{table}

\subsection{Search for the host galaxy}\label{subsec:host}

There is no  host galaxy candidate coincident with the SN location visible in the \ps\ 3$\pi$ images \citep{Chambers2016} or the Dark Energy Camera Legacy Survey \citep[DECaLS;][]{2019AJ....157..168D} and no galaxies with a catalogued spectroscopic redshift in the NASA Extragalactic Database (NED) within a 4.5\arcmin\ radius of \obj\. We combined 328 pre-explosion \ps\ $w$-band (a broad $g+r+i$ composite filter) exposures, with a total effective exposure of 7800s, to create a deep reference stack. The stacked image (Figure~\ref{fig:image}) does not reveal a candidate host galaxy at the location of \obj. Assuming $(w-r) \approx 0$, we estimate a $3\sigma$ limiting magnitude of $r\gtrsim 24.5$ mag (AB). Depending on the adopted distance (see Section\,\ref{subsec:dist}), the upper limit on the host luminosity is $M_r \gtrsim -11$ to $-10$.

The two sources with the least angular offset to the SN location are marked in the $w$-band stack (Figure~\ref{fig:image}). Source 1, offset by 1.2\arcsec\ east and 3.4\arcsec\ south of \obj\ is a red point-like source, a likely foreground star. Source 2, offset by 10.4\arcsec\ west and 3.5\arcsec\ south, is an extended source and was considered a potential host galaxy candidate. A spectrum of the galaxy, obtained from the Inamori Magellan Areal Camera and Spectrograph (IMACS) instrument on the Magellan telescope on 2022 May 20, reveals emission lines at a redshift $z \approx 0.11$, clearly incompatible with any reasonable classification for \obj.

\subsection{Redshift and Distance}\label{subsec:dist}

In the absence of a host galaxy redshift, the redshift was constrained using the spectral template matching tool \textsc{snid} \citep{2007ApJ...666.1024B}. We added the spectra of 2003fg-like events SN 2007if \citep{2010ApJ...713.1073S} and SN 2009dc \citep{2011MNRAS.412.2735T} to the SNID template library that already contained the spectra of SN 2006gz. The top matches on SNID are consistently with SN 2009dc and SN 2006gz for the pre-maximum spectra. The spectra of \obj\ show excellent matches with SN 2009dc and SNID favours a low redshift of $z = 0.020-0.022$. For SN 2006gz, SNID favours a higher redshift of $z = 0.030-0.032$. Fits to normal SNe Ia spectra are inferior around maximum and pre-maximum epochs, although not unreasonable post-maximum. Normal SNe Ia spectra at maximum do not show the prominent and persistent C~{\sc ii} feature exhibited by \obj, that is a signature of 2003fg-like SNe. The normal SN Ia matches require a redshift of $z = 0.030-0.035$, implying  peak absolute magnitudes of $M_g < -20$ for \obj, incompatible with normal SNe Ia. Based on the matches with SN 2006gz and SN 2009dc, we adopt a mean redshift of $z=0.026$, with a range $0.021 < z < 0.031$. 
Assuming a flat Universe with $\Omega_{\rm M}=0.3$ and adopting $H_0 = 70$ km s$^{-1}$ Mpc$^{-1}$ implies a mean luminosity distance of $114$ Mpc with a range $91-136$ Mpc (distance modulus $\mu = 35.28_{-0.47}^{+0.39}$ mag). For the subsequent analysis, we present the properties of \obj\ at this range of distances.

\section{Light Curve and spectral data}

\subsection{Early flux excess}
\label{subsec:excess}

We manually performed forced photometry on all the 30s difference images from ATLAS over the supernova duration and 
before discovery. This forces a PSF fit which is obtained from the individual input images prior to subtraction at the mean position of the source. The forced photometry flux ($f\pm f_{\rm err}$) is calibrated with stars in RefCat2 \citep{2018ApJ...867..105T} in units of micro Jansky ($\mu$Jy). We compute a combined weighted flux ($F$) from the fluxes ($f_i$) measured in the quad of exposures on a nightly basis, using weights dependent on the measurement uncertainties, $w_i = \Delta f_{i,{\rm err}}^{-2}$ : 

\begin{equation}
 F = \frac{\sum_i w_i \times f_i}{\sum_i w_i}\,.
\end{equation}

\begin{figure*}
    \centering
    \includegraphics[width=\linewidth]{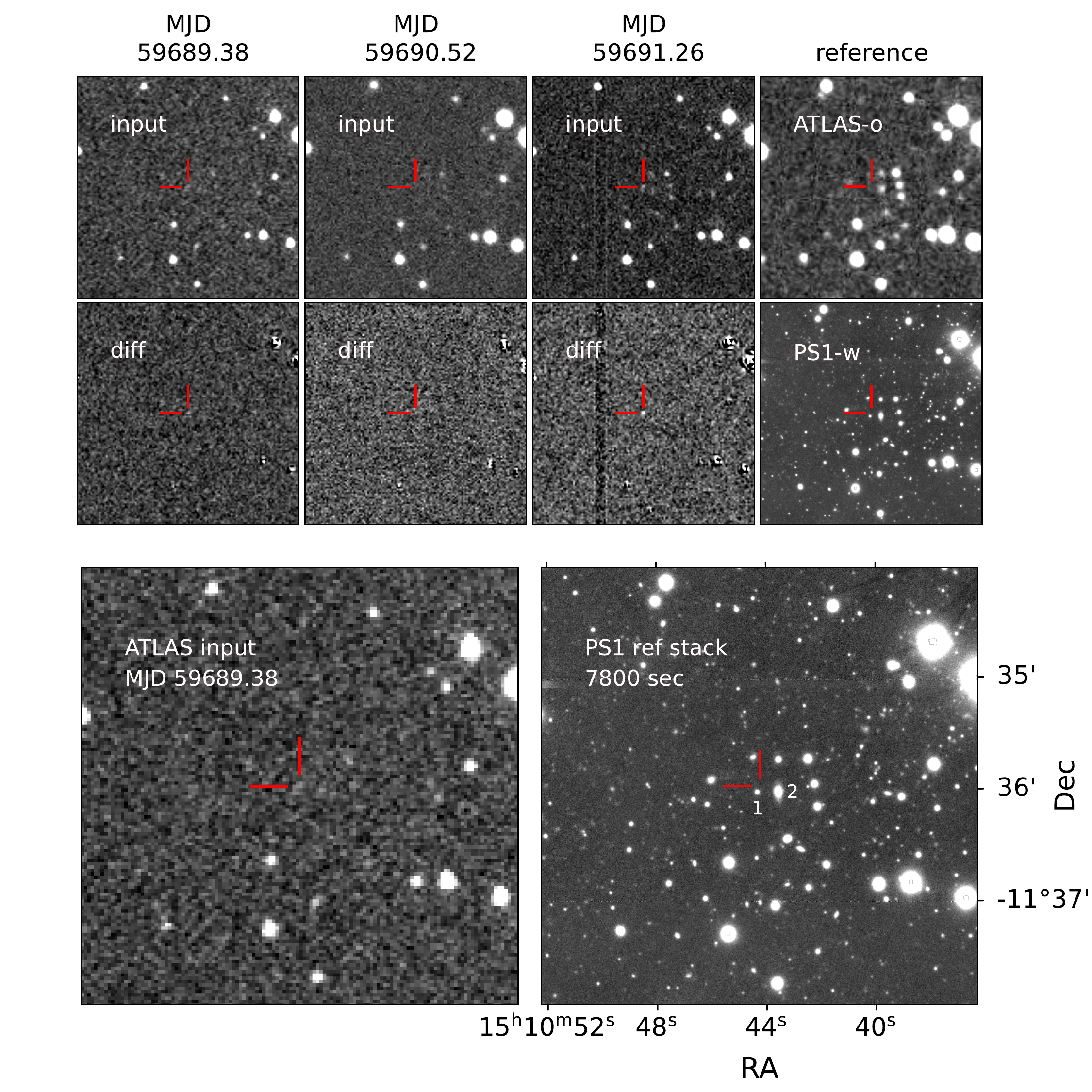}
    \caption{ATLAS and \ps\ images for the field of \obj. Top row (left to right): median-combined ATLAS input images on MJDs 59689.38, 59690.52, 59691.26, and the reference ATLAS template used for the subtraction.
    Middle row: median-combined ATLAS difference images on MJDs 59689.38, 59690.52, 59691.26, and the deep \ps\ $w$-band stack with a effective exposure time of 7800 seconds. Bottom row: median-combined ATLAS input image on MJD 59689.38 and the \ps\ reference stack. The SN position is marked by the cross-hairs.
    Two sources closest to the SN position are marked. Source 1 is a foreground star, and source 2 is a galaxy at $z \approx 0.11$, thus ruled out as a host galaxy candidate for \obj.}
    \label{fig:image}
\end{figure*}

This results in a $3.9\sigma$ detection on MJD=59689.38 (median time of the quad) in $o$-band at $F = 65.7 \pm 16.8\,\mu$Jy or $o=19.36 \pm 0.28$ mag. The stacked measurements show $5\sigma$ detections on MJD 59691.26 and beyond, but we see a non-detection on the intervening epoch of MJD 59690.52, at $F = 23.3 \pm 11.9\,\mu$Jy, corresponding to a 3$\sigma$ upper limit of $o > 20.02$). 
To illustrate the reliability of these combined flux measurements on the individual frames, we also co-added the images. The four individual 30s exposures on each of the MJD epochs 59689.38, 59690.52 and 59691.26 were aligned and combined with median co-addition (Figure~\ref{fig:image}). Visual inspection of the co-added input and difference images confirms the presence of residual flux at the SN location on MJD 59689.38, followed by no obvious detection the next day, on MJD 59690.52. A clear $5\sigma$ detection is recovered on the following day MJD 59691.26, with \obj\ subsequently showing the usual SN-like rise. The detection on MJD 59689.38 and the subsequent non-detection on MJD 59690.52 indicates a short-lived early excess in the light curve of \obj. The timescale of this excess is similar to that of \hvf\ \citep{2021ApJ...923L...8J}. 
Although we do not have as high-cadence observations for \obj\ as those for \hvf, we do have an extra ATLAS observation that further constrains the sharp \hvf\ rise
shown by \cite{2021ApJ...923L...8J}.

\subsection{Multi-band light curves}

\begin{figure*}
    \centering
    \includegraphics[width=\linewidth]{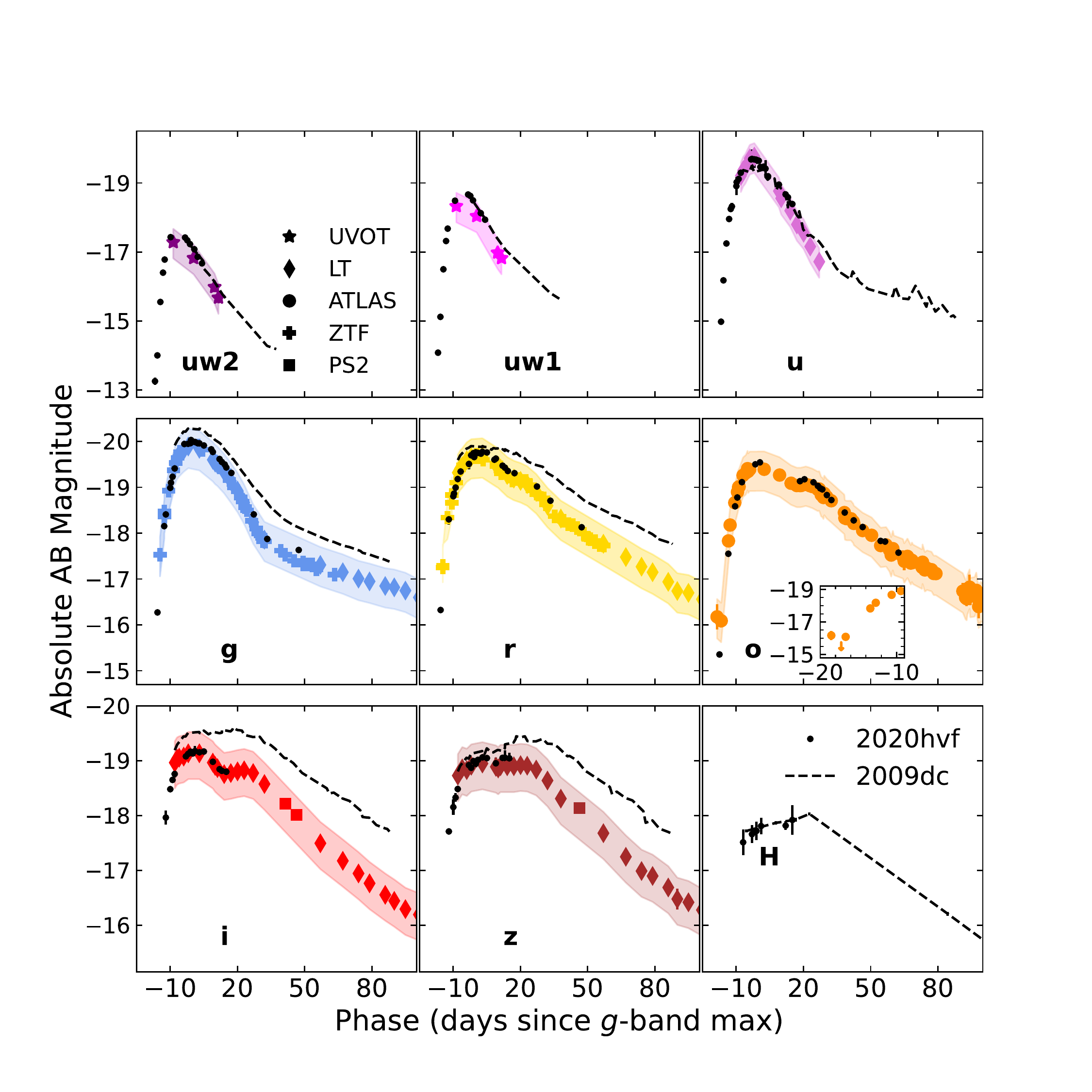}
    \caption{Multi-band light curve evolution of \obj\ (coloured symbols), compared to \hvf\ (this work) and the luminous SN 2009dc \citep{2011MNRAS.410..585S,2011MNRAS.412.2735T}. The absolute magnitudes for \obj\ were computed for the assumed distance of 114 Mpc (corresponding to $z=0.026$). The shaded regions represent the range of absolute magnitudes for a distance range of $91-136$ Mpc. Inset for the $o$-band light curve (middle right panel) shows the early excess detected by ATLAS. All magnitudes are in the AB system.}
    \label{fig:lcplot}
\end{figure*}

The multi-band light curve evolution of \obj\ is shown in Figure~\ref{fig:lcplot}. Also shown are the light curves of \hvf\ (this work) and SN 2009dc \citep{2011MNRAS.410..585S,2011MNRAS.412.2735T}. The observed magnitudes of \obj\ and \hvf\ were corrected for foreground Galactic extinction \citep{2011ApJ...737..103S}, with $R_V = 3.1$ and $E(B-V) = 0.11, \,0.04$, respectively. The $UBVRI$ magnitudes of SN 2009dc were converted to $ugriz$ using the transformations prescribed by \citet{2005AJ....130..873J}.
For a meaningful comparison, we convert the extinction-corrected apparent magnitudes to absolute magnitudes, assuming $\mu = 35.28_{-0.47}^{+0.39}$ for \obj\ (Section~\ref{subsec:dist}). The shaded region in the plot show the range of absolute magnitudes for the range in distance ($91-136$ Mpc) for \obj.

The large uncertainty in distance translates to a large uncertainty on the peak absolute magnitude for \obj, with $M_g^{\rm peak} = -19.89_{+0.47}^{-0.39}$. Even at the lower distance limit of 91 Mpc, the peak luminosity of \obj\ is $M_g^{\rm peak} \gtrsim -19.4$, comparable or more luminous than normal SNe Ia. The $g$-band decline rates for \obj\ and \hvf\ are identical, $\Delta m_{15}(g) = 0.58 \pm 0.05$. The derived $g$-band decline rate for SN 2009dc is also similar within uncertainties, $\Delta m_{15}(g) \approx 0.56 \pm 0.04$. SN 2009dc was very luminous, with a derived $M_g^{\rm peak} \approx -20.3$.

\subsection{Bolometric light curves}\label{subsec:bol}

The quasi-bolometric light curves of \obj\ and \hvf\ were computed from the broadband optical $ugcroiz$ magnitudes using the \texttt{SuperBol} code \citep{2018RNAAS...2d.230N}. The UV-optical-NIR bolometric light curve of SN 2009dc was computed from the published photometry using the same method for consistency. For \hvf, we also compute a UV-Optical-NIR bolometric light curve using the UVOT and LT $H$-band photometry. 

\begin{figure*}
    \centering
    \includegraphics[width=\linewidth]{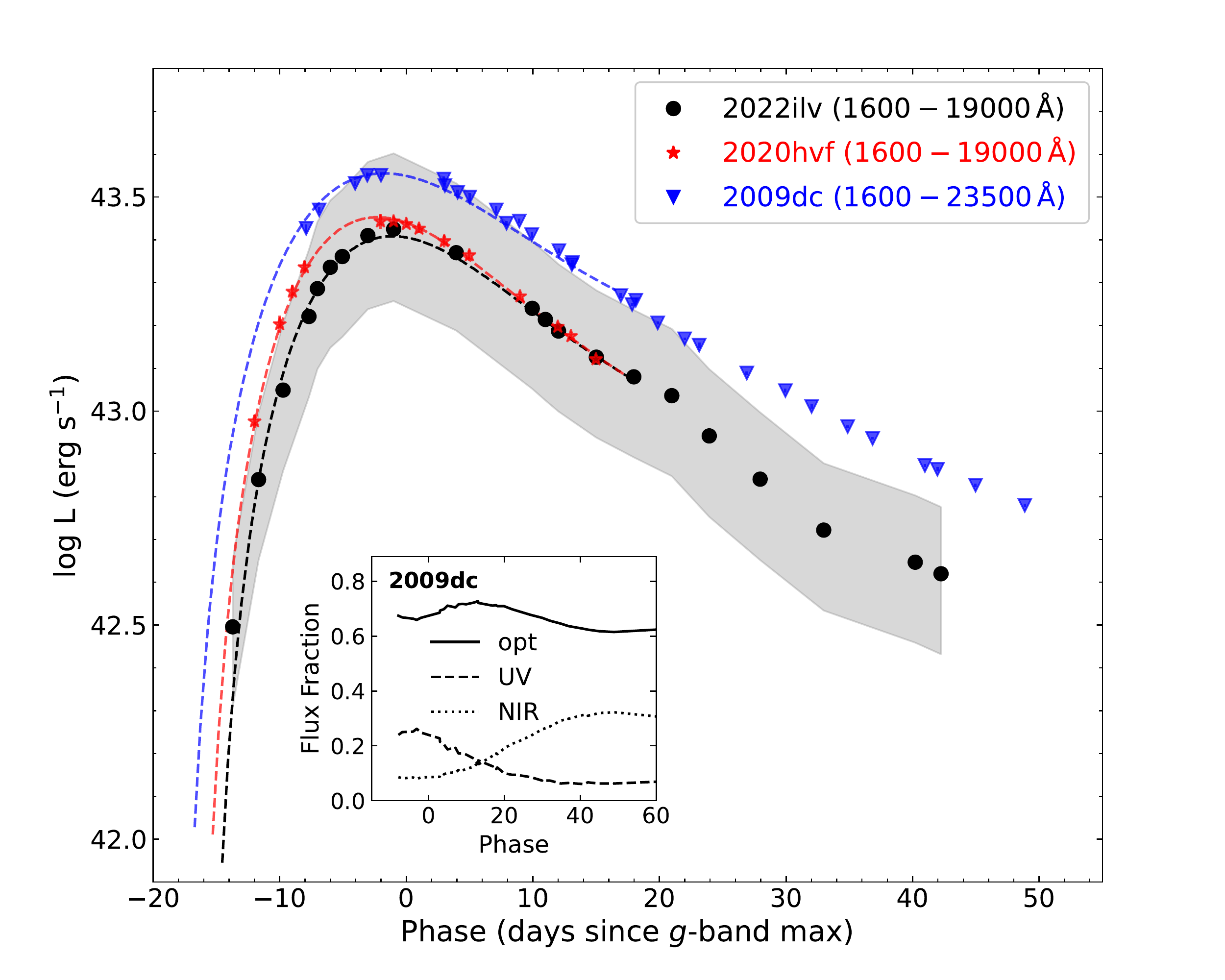}
    \caption{Bolometric light curves of \obj\ ($1600-23500\,$\AA), \hvf\ ($1600-19000\,$\AA) and SN 2009dc ($1600-23500\,$\AA). The shaded region represents the luminosity range for \obj\ given the uncertainty in the distance. Inset shows the fractional flux contribution from optical, UV and NIR bands to the total bolometric flux of SN 2009dc. Dashed lines represent the best-fitting Arnett models for each SN.}
    \label{fig:bolometric}
\end{figure*}

Since \obj\ and SN 2009dc show a similar spectral evolution (Figure~\ref{fig:specplot}), we compute the time-dependent fractional UV ($uvw2-m2-w1$) and NIR ($JHK$) contribution to the bolometric flux for SN 2009dc, and apply those corrections to compute a UV-optical-NIR bolometric light curve for \obj\ assuming a similar fractional contribution in the UV and NIR.
The bolometric light curves of \obj\ ($1600-23500\,$\AA), \hvf\ ($1600-19000\,$\AA) and SN 2009dc ($1600-23500\,$\AA) are shown in Figure~\ref{fig:bolometric}. The shaded region represents the range of luminosity for \obj\ for the distance range. The inset shows the time-dependent fractional flux contribution from the optical ($UBVRI$), UV ($uvw2-m2-w1$) and NIR ($JHK$) to the bolometric light curve of SN 2009dc. We use the Arnett model \citep{1982ApJ...253..785A,2008MNRAS.383.1485V} to estimate explosion parameters from the  bolometric light curves. We fit for $^{56}$Ni mass $M_{\rm Ni}$, ejected mass $M_{\rm ej}$, kinetic energy $E_{\rm k}$ and rise time $t_{\rm r}$  and fix the opacity at $\kappa = 0.1$ cm$^2$ g$^{-1}$. The photospheric velocity $v_{\rm ph}$, from the \SiIIa\ velocity around maximum for each SN was measured to constrain the fits. For \obj, we measure $9000 < v_{\rm ph} < 12000$ km~s$^{-1}$, depending on the adopted redshift, leading to explosion parameters in the ranges $M_{\rm Ni} \sim 0.7-1.5$ \msol, $M_{\rm ej} \sim 1.6-2.3$ \msol, and $E_{51} \sim 0.8-2.0$ foe.

\subsection{Spectral Features}

\begin{figure*}
    \centering
    \includegraphics[width=0.95\linewidth]{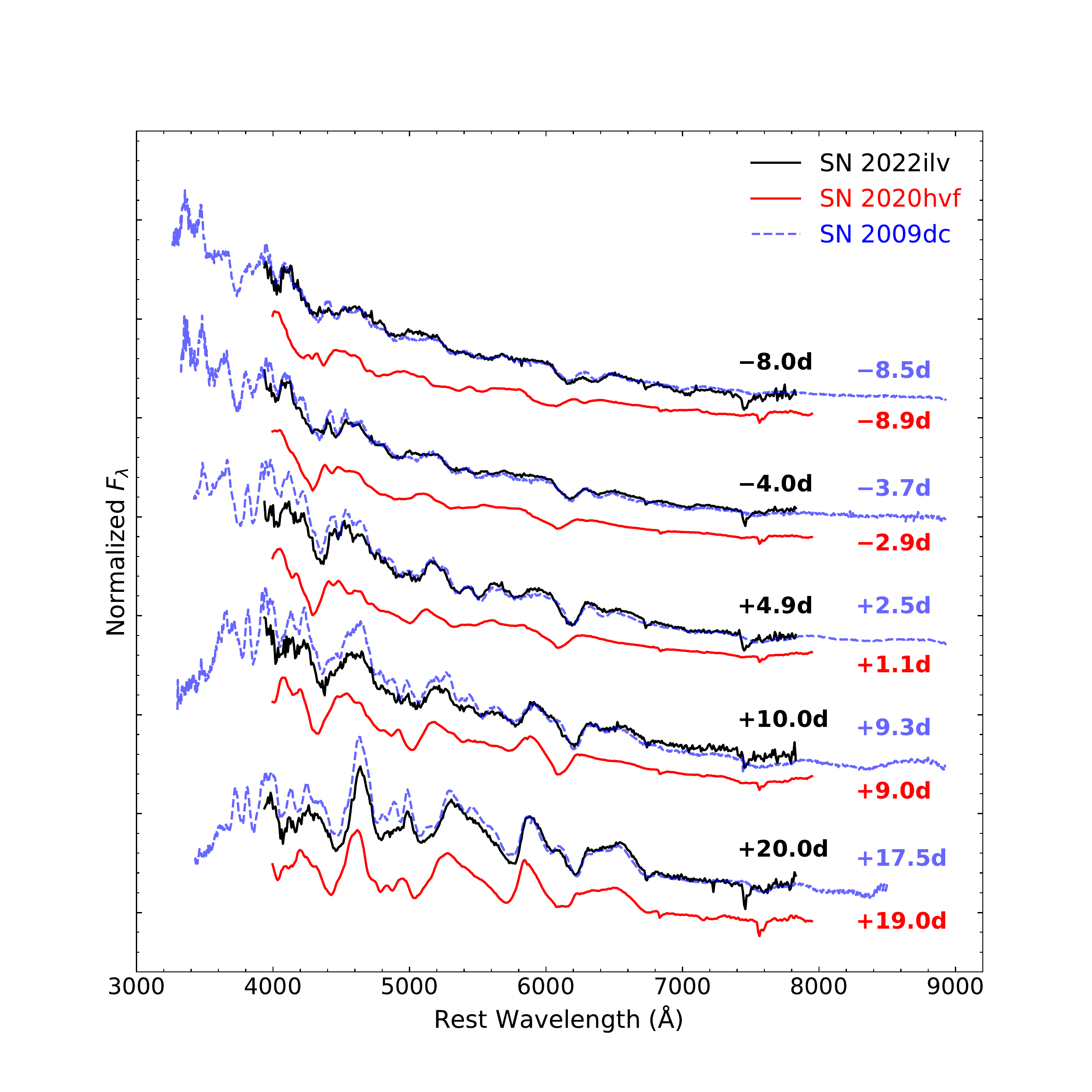}
    \caption{Our LT spectra (with SPRAT) of \obj\ and \hvf\ (this work), plotted with spectra of SN 2009dc \citep{2011MNRAS.412.2735T} at similar epochs for comparison. The spectra of \obj\ and \hvf\ will be made available on the Weizmann Interactive Supernova Data Repository \citep[WISeREP;][]{2012PASP..124..668Y}.}
    \label{fig:specplot}
\end{figure*}

The spectra of \obj\ are almost identical to the luminous, carbon-rich 
SN 2009dc (Figure~\ref{fig:specplot}) for an adopted redshift of  $z = 0.021$. \obj\ shows a prominent C~{\sc ii} $\lambda 6580$ feature in the $-8.9$d spectrum that is comparable in strength to the \SiIIa\ feature, with C~{\sc ii} $\lambda 7234$ also detected. This C~{\sc ii} feature persists until well beyond maximum and is clearly detected in the $+9.1$d spectrum for \obj, as in the case of other 2003fg-like events. 
For the mean adopted redshift of $z = 0.026 \pm 0.005$ for \obj, we deduce a \SiIIa\ velocity of $\sim 11500$ \kms\ at $-8.9$d, slowing down to $\sim 10200$ \kms\ at $+4.0$d, with a systematic uncertainty of $1500$ \kms\ owing to the uncertainty in redshift.

The C~{\sc ii} $\lambda 6580$ feature is also detected in the $-8.9$d spectrum of \hvf. It is weaker than the other two and does not persist beyond the epoch of maximum.
\hvf\ shows very high expansion velocities, particularly in the pre-maximum phase. The broadened and asymmetric \SiIIa\ line profile in \hvf\ likely has contribution from a high velocity ejecta component. Using a Gaussian fit, we deduce a Si~{\sc ii} expansion velocity of $20300 \pm 700$ \kms\ at $-16.9$d, falling to $12800 \pm 600$ \kms\ by $+1.1$d. Broad O~{\sc i} $\lambda 7774$ is also detected in the $-16.9$d spectrum at a comparable velocity of $20200 \pm 800$ \kms.

\section{Analysis of the early flux excess}\label{sec:excess_model}

\begin{figure*}[h]
    \centering
    \includegraphics[width=0.75\linewidth]{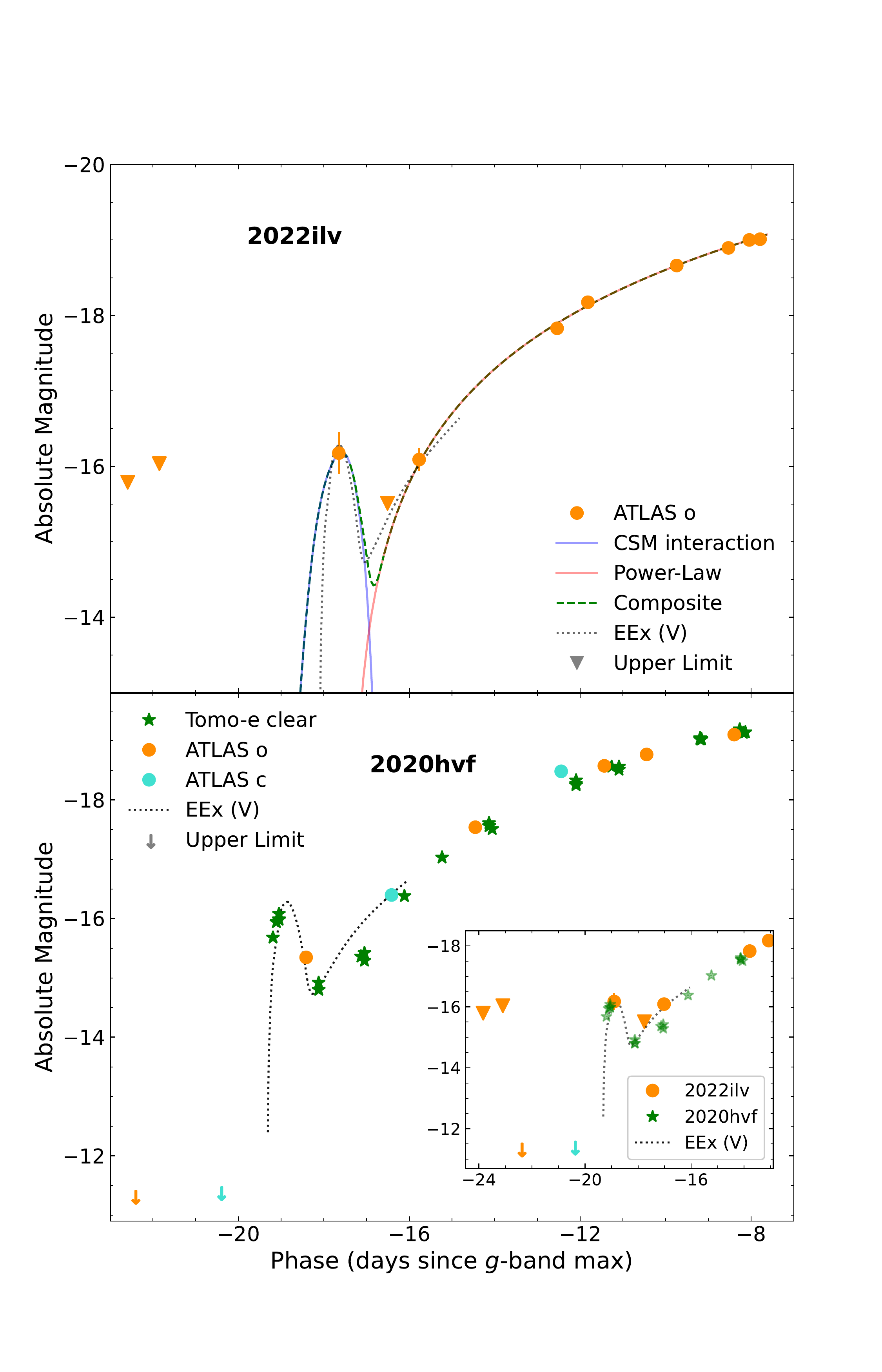}
    \caption{Top panel: ATLAS $o$-band absolute magnitude light curve of \obj, with the synthetic $o$-band \citet{2015ApJ...808L..51P} model  for the interaction-powered early excess, and a power-law fit to the radioactively driven subsequent rise. The composite model, (sum of the two) is shown as a dashed line. A SNEC-based CSM model (EEx) computed by \citet{2021ApJ...923L...8J} to explain the short-lived early excess emission observed in \hvf\ is shown for comparison. Bottom panel: The Tomo-e (clear filter) light curve of \hvf\ from \citet{2021ApJ...923L...8J}, along with our ATLAS photometry that enhances the sampling of the published pulse. Inset: A zoomed-in plot showing a comparison of the early excess feature in \obj\ and \hvf. Since \hvf\ has a longer rise time, the light curve of \obj\ was shifted by $\approx 1$ day to match with \hvf. 
    }
    \label{fig:excess}
\end{figure*}

An early flux excess was detected in high cadence ATLAS observations of the field of \obj\ (Section~\ref{subsec:excess}), the timescale ($\sim 1$\,day) of which is comparable to that detected in  \hvf\ \citep{2021ApJ...923L...8J}. ATLAS also caught \hvf\ during the early excess phase (Figure~\ref{fig:excess}), however the 2-day cadence at the time meant that a positive identification of this feature was not possible. The most promising scenario to explain this pulse-like, short-lived early excess involves interaction of fast-moving SN ejecta with dense, confined CSM \citep[e.g.][]{2016ApJ...826...96P,2018ApJ...865..149J}. The CSM is shock-heated and subsequently cools, powering this initial peak \citep{2015ApJ...808L..51P}. Following the one-zone analytic formulation of \citet{2015ApJ...808L..51P} for the interaction-powered emission, the timescale of the excess depends on the amount of extended CSM as

\begin{equation}\label{eq:tp}
t_{\rm p} \approx \left( \frac{M_{\rm e} \kappa}{v_{\rm e}c} \right) ^{1/2}\,.
\end{equation}

Here, $M_{\rm e}$ is the mass of the extended CSM, $\kappa$ is the electron scattering opacity of the material, and $v_{\rm e}$ is velocity of the shocked CSM. The fraction of kinetic energy from the ejecta that is deposited into the CSM ($E_{\rm e}$) depends on the SN ejecta mass ($M_{\rm ej}$) and kinetic energy ($E_{51}$) as
\begin{equation}
E_{\rm e} \approx 4 \times 10^{49} \, E_{51} \left( \frac{M_{\rm ej}}{\msol} \right)^{-0.7} \left( \frac{M_{\rm e}}{0.01 \, \msol} \right)^{0.7}\,.
\end{equation}

The peak luminosity for the emission is given by
\begin{equation}\label{eq:Lp}
L_{\rm p} \approx \frac{E_{\rm e} R_{\rm e}}{v_{\rm e}{t_{\rm p}^2}}\,.
\end{equation}
Here, $R_{\rm e}$ is the initial extent, or inner radius of the extended CSM. The radius of the extended material as a function of time is $R(t) = R_{\rm e} + v_{\rm e} t$. The time-dependent interaction-powered luminosity is thus denoted as 
\begin{equation}\label{eq:Lt}
L(t) = \frac{t_{\rm e} E_{\rm e}}{t_{\rm p}^2} \exp \left[ - \frac{t(t+2t_{\rm e})}{2t_{\rm p}^2} \right],
\end{equation}
where $t_{\rm e}$ is the characteristic expansion timescale, $t_{\rm e} = R_{\rm e}/v_{\rm e}$. We note that the model is quite degenerate, and the interaction-powered luminosity (Equation~\ref{eq:Lt}) is sensitive to the choice of several parameters such as $M_{\rm e}$, $\kappa$ and $E_{51}$, that in turn depends on ejecta mass $M_{\rm ej}$ and the bulk photospheric velocity for the ejecta, $v_{\rm ph}$. Equation~\ref{eq:tp} suggests that the timescale of the emission depends primarily on the CSM mass $M_{\rm e}$. The duration of the pulse therefore provides an important constraint on the amount of extended material around the WD progenitor. The peak luminosity of the pulse, on the other hand, depends mainly on the radial extent of the CSM rather than its mass (Equation~\ref{eq:Lp}).

Using the constraints from modeling the bolometric light curve of \obj\ (Section~\ref{subsec:bol}), we fix $M_{\rm ej} \sim 2$~\msol\ and $v_{\rm ph} \sim 10000$ \kms. Given the degeneracy in the model and the uncertainty on distance, we expect these parameters to be uncertain by a factor of $\sim 2$. The electron scattering opacity was set at $\kappa = 0.2$ cm$^2$ g$^{-1}$ assuming hydrogen-poor CSM \citep{2015ApJ...808L..51P}. For extended material present at a distance of $R_{e} \sim 10^{13}$ cm, the resulting ejecta-CSM interaction and subsequent shock-cooling is expected to produce a luminous excess mostly in the optical/UV \citep{2013ApJ...772....1R}. The luminosity of the excess emission detected by ATLAS, at $M_o \sim -16.2$ mag, is consistent with CSM present at $R_{\rm e} \sim 10^{13}$ cm. Finally, the $\sim 1$ day timescale of the excess emission indicates a small amount of CSM. We find that a CSM mass of $M_{\rm e} \sim 10^{-3}$ \msol\ provides a reasonable fit to the observed excess. Assuming blackbody emission, we estimated a temperature for the shock-heated CSM, and model light curves in different bands were computed from synthetic photometry using \texttt{synphot} \citep{2018ascl.soft11001S}. 

The post-excess, rising $o$-band light curve of \obj\  was fit with a simple power-law \citep{2015MNRAS.446.3895F}, as 
\begin{equation}
    F = \alpha (t-t_0)^n\,.
\end{equation}

The constant  $\alpha$,  time of explosion $t_0$, and the exponent $n$ were varied as free parameters and the best-fit values are $n = 1.55 \pm 0.18$, $t_0 = 59689.66 \pm 0.69$ (corresponding to $-17.8 \pm 0.7\,$ rest-frame days relative to $g$-band maximum). A composite model for the early light curve was computed by summing the flux from the ejecta-CSM interaction model of \citet{2015ApJ...808L..51P}  and the power-law model of the radioactively driven rise (see Figure~\ref{fig:excess}). 
Also shown is the synthetic $V$-band light curve computed by \citet{2021ApJ...923L...8J} to explain the observed early excess in \hvf. The \citet{2021ApJ...923L...8J} model involves a more sophisticated treatment for the density profile of the extended CSM formulated by \citet{2016ApJ...826...96P}, and synthetic light curves were computed using the SuperNova Explosion Code \citep[\texttt{SNEC};][]{2015ApJ...814...63M}. 

The \citet{2021ApJ...923L...8J} model for the early excess of \hvf\ provides a very reasonable fit, without any scaling in time or luminosity axes, to the early excess and subsequent rise for \obj. The timescale of our \citet{2015ApJ...808L..51P} model is comparable to that of the \citet{2021ApJ...923L...8J} EEx model. However, we note that \citet{2021ApJ...923L...8J} estimate a CSM mass of $10^{-2}$ \msol, roughly an order of magnitude higher than our estimate based on the simpler analytic approach. 

The inset in Figure~\ref{fig:excess} shows the early excess feature in the two events \obj\ and \hvf. Although the features are comparable in terms of luminosity and timescale, we are limited in constraining the CSM properties due to the uncertainty in the distance for \obj\ and the cadence of the observations. Although the uncertainty in distance yields large uncertainties on the explosion parameters for \obj, the mean $M_{\rm ej}$ of $\sim 2$ \msol\ is similar to that inferred for \hvf. Assuming the CSM opacity $\kappa$ and timescale $t_p$ of the two events is similar, Equation~\ref{eq:tp} suggests that the higher photospheric velocity of \hvf\ around peak ($\sim 13000$ \kms, in contrast to $\sim 10000$ \kms\ for \obj) would imply a higher CSM mass for \hvf\ compared to \obj\ by a factor of $\sim 1.3$. High-cadence, multi-band observations of these early excess features in more such Ia-SC events will reveal further insights into the properties of the CSM around their WD progenitors.

\section{Discussion and Conclusions}

The early excess detected in these two peculiar and luminous SNe Ia (\hvf\ and \obj) was tentatively also detected in 2003fg-like events ASASSN-15pz \citep{2019ApJ...880...35C} and LSQ12gpw \citep{2018ApJ...865..149J}. This feature could be common within the family of carbon-rich, luminous SNe Ia. The short timescale of the excess, low intrinsic volumetric rate of these events, and limitations in cadence and sensitivity of surveys are factors that may have curtailed previous discoveries. Rapid real-time detection and followup of this early excess feature in more such events will enhance our understanding of the nature of the progenitor.

A lingering question for 2003fg-like events is whether the high luminosities of the main peaks are purely radioactively driven, or if interaction with surrounding material boosts the (otherwise normal) luminosity. Super-$M_{\rm Ch}$ WD progenitors have been invoked for many 2003fg-like events \citep{2017hsn..book..317T} to explain their extreme properties. 
Such super-$M_{\rm Ch}$ models succeed in reproducing the high luminosity and broad light curve shape. 
Double WD merger simulations show that tidal stripping of the secondary can eject $\sim 10^{-4} - 10^{-2}$ \msol\ of material \citep{2013ApJ...772....1R,2014MNRAS.438...14D}, forming a carbon/oxygen-rich CSM. If the time lag between merger and explosion is a few hours ($\sim 10^4$ seconds), the CSM would be placed at $\sim 10^{13}$ cm from the WD \citep{2013ApJ...772....1R}, consistent with our estimates from modelling the early excess for \obj\ (Section~\ref{sec:excess_model}). However, the super-$M_{\rm Ch}$ scenario has some shortcomings. Nebular spectra of SN 2009dc and SN 2012dn (among other 2003fg-like SNe) show clear evidence of a low ionization state \citep{2013MNRAS.432.3117T,2019MNRAS.488.5473T}, contrary to what is expected for a large amount of heating from a high $^{56}$Ni mass.

A plausible alternative to the super-$M_{\rm Ch}$ scenario is a normal $M_{\rm Ch}$ explosion occuring within a dense, carbon and oxygen-rich envelope \citep[e.g.][]{2012MNRAS.427.2057H,2013MNRAS.432.3117T,2016MNRAS.463.2972N}. The empirical correlation between Si~{\sc ii} velocity and strength of the C~{\sc ii} feature in a sample of 2003fg-like events \citep{2021ApJ...922..205A} supports this idea. It may explain the prominent and persistent C~{\sc ii} features observed owing to the large amount of unburnt material. SN 2020esm \citep{2022ApJ...927...78D} shows pre-maximum spectra almost entirely dominated by unburnt carbon and oxygen. The interaction of SN ejecta with the surrounding material provides enhanced luminosity, and the reverse shock would decelerate the ejecta \citep{2016MNRAS.463.2972N}, potentially explaining the combination of high luminosity and low velocity. \citet{2016MNRAS.463.2972N} estimated a carbon/oxygen-rich CSM mass of $\sim 0.6$ \msol\ for SN 2009dc through detailed hydrodynamic simulations. 
While their models reproduced the light curve 
shape of of SN 2009dc remarkably well, it had too low a luminosity by a factor of 0.2\,dex. The early excess observed in \obj\ requires $M_{\rm e} \sim 10^{-3}$ \msol\ within a radius 
$R_{\rm e} \sim 10^{13}$\,cm, implying a density of order $\sim10^{-11}$ g\,cm$^{-3}$. 
The  \citet{2016MNRAS.463.2972N} model requires a much larger mass of $\sim 0.6$ \msol\ with a power law density profile between $\sim10^{-9}$ g\,cm$^{-3}$ and $\sim10^{-10}$ g\,cm$^{-3}$ out to $R_{\rm e} \sim 1.4\times10^{14}$\,cm. If CSM contributes to both the early excess and the main luminosity peak, it would likely require two separate components or a structured CSM profile.

If such a significant amount of carbon/oxygen-rich CSM is the explanation for either the pulse or main peak,  then its origin remains mysterious. This could be a natural outcome of the double degenerate (DD) scenario, wherein the disruption and subsequent accretion of the secondary WD leads to the formation of a hot, spherical CO envelope and a centrifugally supported disk \citep{2007MNRAS.380..933Y}. Simulations have shown that the outcome is sensitive to whether the explosion occurs promptly after the merger \citep{2014ApJ...785..105M}, or after a lag \citep{2014ApJ...788...75R}. 
 \cite{2020ApJ...900..140H} and \cite{2021ApJ...922..205A} have suggested an alternative of a CO WD merging with the degenerate core of an asymptotic giant branch (AGB) star, or the core-degenerate scenario \citep{2011MNRAS.417.1466K}. This scenario predicts a significant X-ray luminosity due to the interaction, that should be detectable in nearby events. The lack of narrow emission features in observed spectra of 2003fg-like events is a significant shortcoming of all interaction-based scenarios. 
 Photometric, spectroscopic and polarimetric observations of more such events, in conjunction with modelling efforts and binary population synthesis calculations, will be necessary for unraveling the progenitor puzzle.

The luminous family of 2003fg-like SNe Ia are known to occur preferentially either in low luminosity hosts, or remote locations in luminous hosts \citep{2017hsn..book..317T}. 
The limit of $M_r \gtrsim -11$ for the host of \obj\  (Section~\ref{subsec:host}) is consistent with these findings. The luminous lenticular galaxy NGC 5872 \citep[$z=0.024556$;][]{2015ApJS..218...10V} is at the right redshift, and is offset by 7.7\arcmin\ from the location of \obj\, or a projected radial separation of $\sim 250$\,kpc. This could imply the host is a dwarf galaxy satellite of this more luminous and massive galaxy. The striking preference for environments that are likely low-metallicity should be considered for progenitor modelling and scenarios.

\begin{acknowledgments}

We acknowledge funding from UKRI STFC grants ST/T000198/1 and Lasair funding through  ST/N002512/1 and ST/S006109/1. The Pan-STARRS telescopes are supported by  NASA Grants NNX12AR65G and NNX14AM74G. ATLAS is primarily funded through NASA grants NN12AR55G, 80NSSC18K0284, and 80NSSC18K1575. The ATLAS science products are provided by the University of Hawaii, QUB, STScI, SAAO and Millennium Institute of Astrophysics in Chile. JG is supported by the European Research Council (ERC) under the Consolidator grant BHianca (Grant agreement No. 101002761). MN is supported by the ERC under the European Union’s Horizon 2020 research and innovation program (grant agreement No.~948381) and by a Fellowship from the Alan Turing Institute. KM is funded by the EU H2020 ERC (grant No. 758638). We thank the anonymous referee for helpful suggestions.

\end{acknowledgments}

%

\facilities{ATLAS, Pan-STARRS, ZTF, LT, UVOT}


\software{astropy \citep{2022ApJ...935..167A}, SuperBol \citep{2018RNAAS...2d.230N}, synphot \citep{2018ascl.soft11001S}}





\bibliography{references}
\bibliographystyle{aasjournal}



\end{document}